\documentclass[preprint]{aastex}
\shorttitle{Local Group center}
\shortauthors{A. B. Whiting}
\begin{document}
\title{The kinematical center and mass profile of the Local Group}
\author{Alan B. Whiting}
\affil{University of Birmingham,
Edgbaston Road, Birmingham B15 2TT, UK}

\begin{abstract}
Abandoning the assumption that light traces mass, I seek the
location of the centre of the Local Group of galaxies based solely on
kinematic data and the plausible assumption of infall.  
The available set of positions and radial velocities
is shown to be a misleading indicator of Local Group motions,
giving a direction to the centre offset from the true one; statistical
techniques of moderate sophistication do not catch the offset.
Corrected calculations show the centre to lie in the direction to
M31 within the uncertainty of the method, a few degrees.  
The distance to the centre
is not well determined, lying about 0.5 Mpc from the Milky Way.
The pattern of observed (galactocentric) radial velocities excludes
both dynamically important `orphan haloes' and any extended dark matter 
halo for the Group as a whole, and shows the Group to have formed
from a much more extended volume than it presently occupies.  
Kinematics alone indicates that the mass
of the Group is concentrated effectively in M31 and the Milky Way.

\end{abstract}


\section{The Local Group: mass and motion}

The Local Group of galaxies, the concentration of perhaps four dozen
galaxies around the Milky Way and M31, has proved useful for many years
as a dynamical and cosmological laboratory.  Not only does it contain
those objects that may be studied in greatest detail (due to their
proximity), it was for long the only such structure whose
shape in three dimensions was well-understood.

Perhaps the first important dynamical study
based on the Local Group was
the venerable timing argument of \citet{KW59}.  Making the
assumptions that the only significant mass in the Group
resided in the Milky Way and M31 and that each of these might be
treated as a point particle all the way
back to the Big Bang, the
resulting estimate of mass for the whole Local Group was an early
indication of the presence of large amounts of dark matter.
The argument was developed by \citet{LB81} to include other
galaxies in the Group, assumed to be massless and on radial trajectories
directed toward or away from the barycentre.  In spite of the
simplistic picture of the Group that is assumed, the timing
argument remains a useful point of reference and updated treatments
and versions of it may be found in, for example,
\citet{LB99}, \citet{W99} and \citet{BT08} pp. 150, 268.

A much more sophisticated look at Local Group dynamics is found in
the Least Action calculations as developed, for instance, in
\citet{PPST01} and found most recently in \citet{PTS11}.  Here the trajectories
of the various galaxies are subject to only reasonable restrictions.
The assumptions are retained, however, of the Milky Way and M31
containing most of the Group's mass (though their masses and those
of others are allowed to vary somewhat between solutions) and of galaxies
retaining their identities back to a very
early time. 

There is no question that galaxies have mass or that the Milky Way
and M31 have rather a lot of it.  But the assumption of a close relation
between mass and light remains an assumption on scales of, say,
100kpc to several Mpc, and therefore should be examined if possible.
This is especially so when there are indications pointing toward
its modification.  \citet{DL95} found significant
discrepancies between the parameters of n-body simulations and the
corresponding quantities calculated by treating the simulations as
a least-action problem, discrepancies they attributed to the
presence of `orphan' dark-matter haloes containing no galaxies.
More recently, \citet{W05} discovered that the kinematics of nearby
galaxy groups does not match that expected qualitatively, a
discrepancy that \citet{MYH07} attribute again to `orphan' dark-matter
haloes.   Along similar lines to \citet{DL95}, \citet{LW08}
work out a correction for the timing argument based on more recent
n-body work.  Motivated by the picture of dark matter being far more
extended than visible galaxies, in their examination of the fate of the 
Local Group, \citet{CL08} assume a diffuse intergalactic medium containg as
much mass as the two major galaxies.

The purpose of the present study is to discard the assumption of light
tracing mass and see how far we can go without it.  In
particular, what can the motions alone
of galaxies within the Local Group tell us
about its mass distribution?  {\bf We will use galaxies as tracers
of the velocity field, and (assuming velocities to be due to gravity)
use the velocity field to infer the mass distribution.  In particular
we are looking to see whether the inferred mass matches the location
of visible galaxies, or perhaps indicates the existence of dynamically
important `orphan haloes' or an extended dark-matter halo.}

\subsection{The picture}

We will assume that there is a mass concentration associated with the
Local Group, such as to produce an identifiable centre.  
Next, on
consideration of crossing times, the Group 
(at least beyond a certain distance from the centre) must still be in the
process of infall, and it is plausible that
speeds of infall will be larger closer to the
centre.  Outside a certain radius galaxies will physically
moving away, though more slowly than in the general Hubble flow.
The situation is sketched in Fig. 1.

{\bf This assumption of a mass concentration and radial infall is
plausible, but must be checked when possible.  Some possible signs indicating
failure are mentioned below.}

\begin{figure}
\plotone{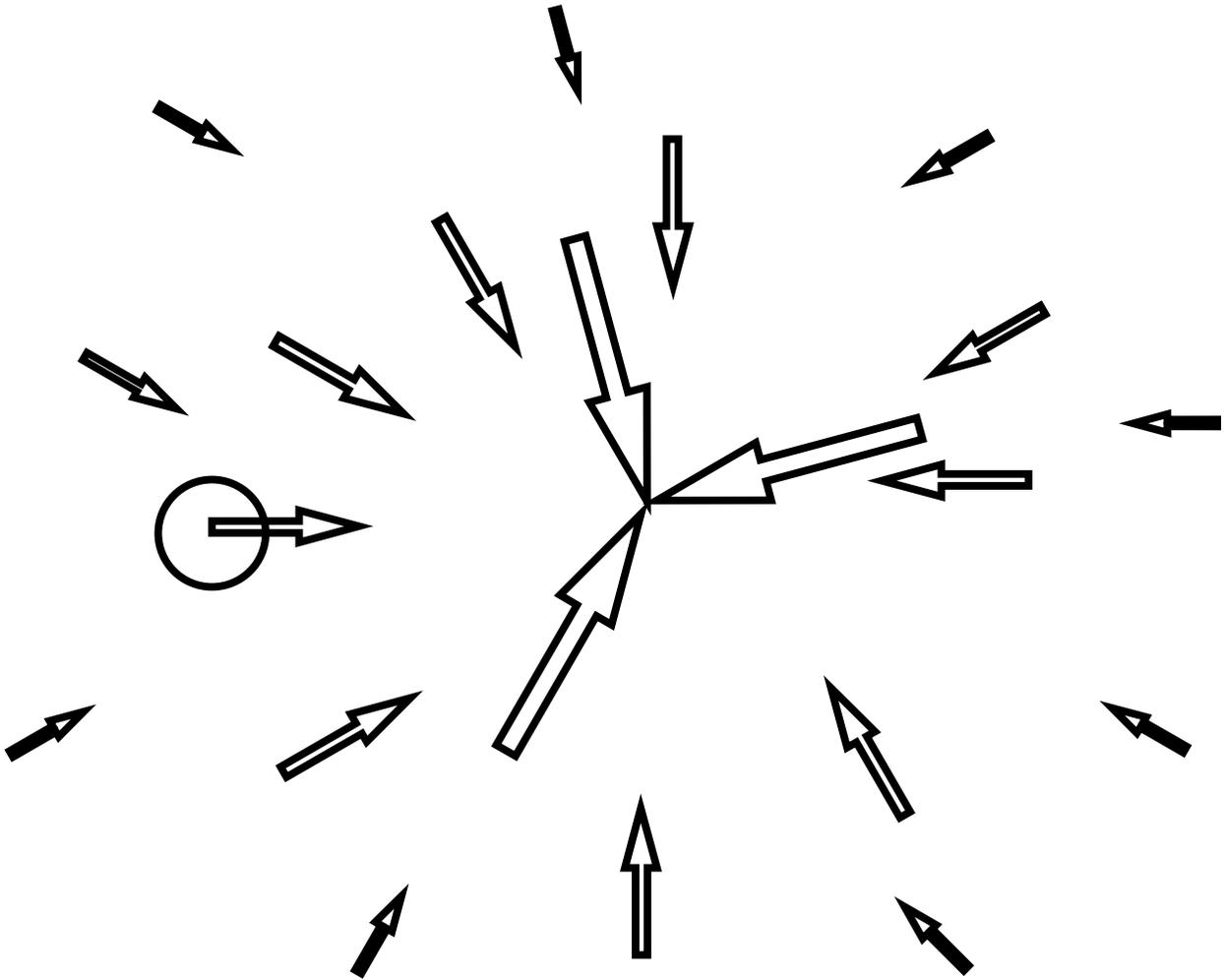}
\caption{A sketch of the assumed dynamical situation of the Local
Group.  Galaxies are falling in to the centre (a few, not
shown, may have passed it and are coming out again).  Their
speeds are larger as they are closer to the centre.  Outside
the region shown galaxies will begin to move away as they
join the Hubble flow.  As seen from one of the galaxies (marked
with a circle), radial
velocities will be of smallest magnitude
at roughly right angles to the
direction to the centre due to geometric effects.  
On the line to the centre,
galaxies on the same side as the observer will show
positive radial velocities; beyond the barycentre, galaxies will
show negative radial velocities.  The greatest radial
speeds will be found on this line.}
\label{picture}
\end{figure}

As observed from a position outside the centre and participating in
the infall, radial velocities will be positive toward the centre and away
it; very negative on the far side; and smaller in magnitude looking off this
axis, for purely geometric reasons.  Close to the
centre the picture may be complicated by galaxies that have fallen in
from the far side and are on their way out again, but in any case the
maximum observed radial speeds (positive or negative) will be in the
direction of the centre and directly away from it.
Within this picture, then, the direction of maximum magnitude
of radial velocities will point toward the centre of the Group.
The distance to the centre will be signaled by a sudden shift from
positive to negative radial velocities or by a mixture of positive and
negative velocities in some region.

The picture need not be exact to be useful.  In particular, the centre itself
may be unoccupied by anything observable, and the velocity field may have
holes in it at any point.  Small-scale deviations and uncorrelated
motions should average out, though in a sparsely-sampled Group they might
not do so as well as we would wish.  

Conversely,  a direction of maximum radial speeds can be calculated
for any situation, even one in which there is no identifiable 
centre, or in which the field of velocities does not have an overall pattern.
We could detect the latter cases by taking various subsets of the
observations and performing a calculation: the direction would fluctuate
strongly depending on the particular galaxies included.
In the case of the Local Group, if
mass indeed follows light we expect the centre to lie in the direction
of M31.

An offcentre mass concentration within the Group
with galaxies around it will show a
large radial-velocity signal not associated with the Group as a whole; how
this is dealt with in the case of M31 and the Milky Way is set
out below.

For this calculation we must correct heliocentric
radial velocities for the motion of the Sun around the Galaxy, a correction
that is not as accurately known as one might like, but we may use
the resulting galactocentric radial velocities with no further correction.

\section{Finding the vertex}

\subsection{The sample}

In attempting an analysis of the dynamics of the Local Group one must
first decide where its limits are.  Including galaxies in the neighboring
galaxy groups risks distorting one's conclusions by the actions
of masses there, so
no objects beyond 1.5Mpc should be used; and those close to the border
should be scrutinised for possible disturbance.  For our purposes
a near limit needs to be chosen also.  Including the many satellite
galaxies of the Milky Way and M31 would show that those dominant
galaxies have significant mass concentrations, which is not in question.
But we seek kinematic clues to mass distribution on a larger scale, so
satellites must be excluded.  How far out the influence
of the bright galaxies dominates kinematics is not clear;
\citet{McC09} discovered evidence that M33, over 200kpc from M31, may have
been influenced by the latter significantly.  For the following calculations I
adopt 200kpc as a standard cutoff distance, though I will occasionally
explore others. 

\begin{figure}
\plotone{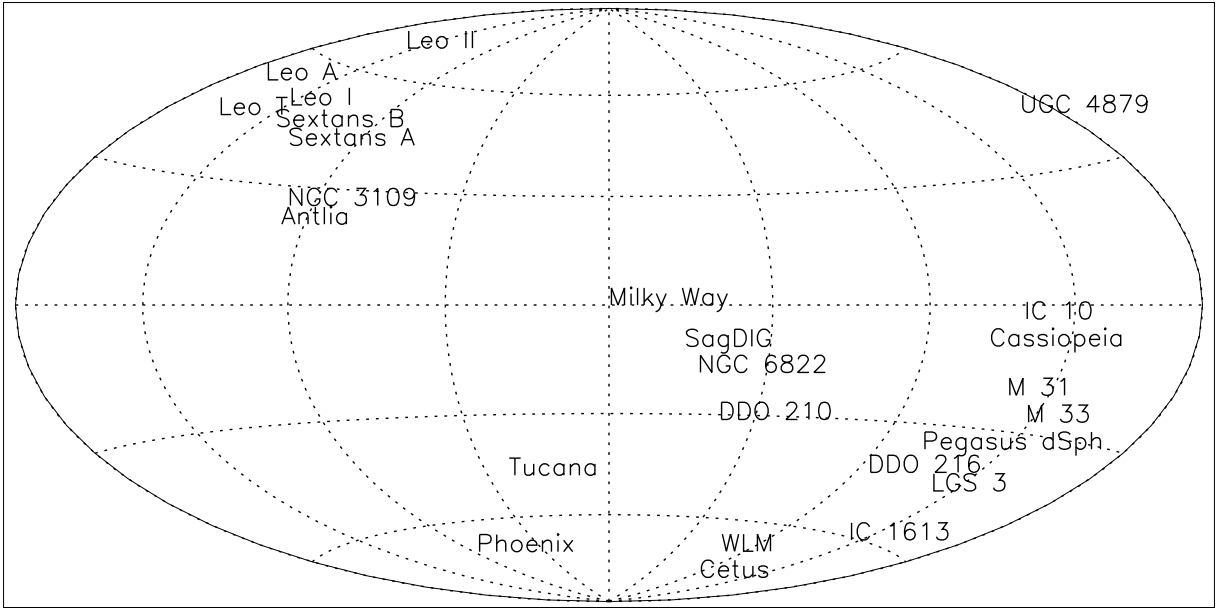}
\caption{The distribution of Local Group galaxies on the sky, as
seen from Earth in galactic coordinates.  Only those more than
200kpc away from both the Milky Way and Andromeda are plotted; some
names have been shifted very slightly for legibility.  The concentration
in upper left and lower right quadrants is evident.}
\label{nameplot}
\end{figure}

Data for galaxies used in the kinematic centre calculations were taken from
the NASA Extragalactic Database (NED)\footnote{This research has made use
of the NASA/IPAC Extragalactic Database (NED) which is operated by the
Jet Propulsion Laboratory, California Institute of Technology, under contract
with the National Aeronautics and Space Administration.}, with the
particular help of Ian Steers.  All galaxies within 1.5Mpc of either
the Milky Way or M31 were initially selected, and those objects that
are clearly part of the Galactic or Andromeda systems (such as the
Sagittarius galaxy, now being cannibalised, and a few globular clusters)
removed.  Objects without a radial velocity could not be used and were
dropped.  Some of the averaged distances listed in NED were adjusted to
give greater weight to newer and more accurate results.  From the
list of 44 galaxies thus obtained, an initial cut to remove satellite
galaxies within 100kpc of M31 or the Milky Way brought the total down
to 33; most of the calculations were done on a final set of 25 using a
200kpc cut.  The galaxies and relevant data are listed in Table~\ref{data}.
The data here used for the 25-galaxy set are almost identical with
the slightly larger sample of \citet{PTS11}.  The distribution 
of the final 25-galaxy sample on the sky is
shown in Fig.~\ref{nameplot}.

\clearpage

\begin{deluxetable}{lcccc}
\tablecaption{Galaxies used in Local Group center calculations.\label{data}}
\tablewidth{0pt}
\tablehead{
\colhead{Name} & \colhead{l} & \colhead{b} & \colhead{$\mu$} &
\colhead{rv}
}
\startdata
Milky Way & 0 & 0 & 0.05 & 2 \\
LMC* & 280.5 & -32.9 & $18.48 \pm 0.1$ & $278 \pm 2$ \\
SMC* & 302.8 & -44.3 & $18.85 \pm 0.1$ & $158 \pm 4$ \\
UMi* & 105.0 & 44.8 & $19.4 \pm 0.1$ & $-247 \pm 1$ \\
Draco* & 86.4 & 34.7 & $19.6 \pm 0.2$ & $-292 \pm 1$ \\
Sextans* & 243.5 & 42.3 & $19.8 \pm 0.1$ & $224 \pm 2$ \\
Sculptor* & 287.5 & -83.2 & $19.64 \pm 0.05$ & $110 \pm 1$ \\
Carina** & 260.1 & -22.2 & $20.05 \pm 0.05$ & $229 \pm 60$ \\
Fornax** & 237.1 & -65.7 & $20.65 \pm 0.1$ & $53 \pm 1$ \\
Leo I & 226.0 & 49.1 & $22.0 \pm 0.5$ & $ 285 \pm 2$ \\
Leo II & 220.2 & 67.2 & $21.65 \pm 0.05$ & $79 \pm 1$ \\
Phoenix & 272.2 & -68.9 & $23.04 \pm 0.05$ & $56 \pm 29$ \\
NGC 6822 & 25.3 & -18.4 & $23.45 \pm 0.05$ & $-57 \pm 2$ \\
M31 & 121.2 & -21.6 & $24.45 \pm 0.05$ & $-300 \pm 4 $ \\
M32* & 121.2 & -22.0 & $24.4 \pm 0.1$ & $-200 \pm 6 $ \\
NGC 205* & 120.7 & -21.1 & $24.5 \pm 0.1$ & $-241 \pm 3 $ \\
And I* & 121.7 & -24.8 & $24.47 \pm 0.05$ & $-368 \pm 11$ \\
And III* & 119.4 & -26.3 & $24.38 \pm 0.05$ & $-351 \pm 9$ \\
NGC 147** & 119.8 & -14.3 & $24.2 \pm 0.05$ & $-193 \pm 3$ \\
And V** & 126.2 & -15.1 & $24.52 \pm 0.08$ & $-403 \pm 4$ \\
And II** & 128.9 & -29.2 & $24.03 \pm 0.1$ & $-188 \pm 3$ \\
NGC 185** & 120.8 & -14.5 & $24.0 \pm 0.1$ & $-202 \pm 3$ \\
Cassiopeia & 109.5 & -10.0 & $24.4 \pm 0.1$ & $-307 $ \\
IC 10 & 119.0 & -3.3 & $24.57 \pm 0.05$ & $-348 \pm 1$ \\
Pegasus dSph & 106.0 & -36.3 & $24.53 \pm 0.06$ & $-354 \pm 3$ \\
LGS 3 & 126.8 & -40.9 & $24.4 \pm 0.1$ & $-287$ \\
DDO 216 & 94.8 & -43.6 & $24.9 \pm 0.2$ & $-183$ \\
Leo T & 214.9 & 43.7 & $23.09 \pm 0.05$ & $35$ \\
Leo V** & 261.9 & 58.5 & $21.23 \pm 0.02$ & $173 \pm 3$ \\
And XIV** & 123.0 & -33.2 & $24.33 \pm 0.33$ & $-481 \pm 1$ \\
And IX* & 123.2 & -19.7 & $24.45 \pm 0.04$ & $-216$ \\
UGC 4879 & 164.7 & 42.9 & $25.22 \pm 0.2$ & $-70 \pm 15$ \\
IC 1613 & 129.7 & -60.6 & $ 24.4 \pm 0.1$ & $-234 \pm 1$ \\
Cetus & 101.5 & -72.9 & $24.44 \pm 0.03$ & $-87$ \\
Leo A & 196.9 & 52.4 & $24.5 \pm 0.1$ & $24$ \\
WLM & 75.9 & -73.6 & $24.9 \pm 0.1$ & $-122 \pm 2$ \\
Tucana & 322.9 & -47.4 & $24.72 \pm 0.05$ & $194 \pm 4$ \\
DDO 210 & 34.0 & -31.3 & $24.97 \pm 0.1$ & $-141 \pm 2$ \\
SagDIG &21.1 & -16.3 & $25.15 \pm 0.5$ & $-79 \pm 1$ \\
NGC 3109 & 262.1 & 23.1 & $25.54 \pm 0.05$ & $403 \pm 1$ \\
Antlia & 263.1 & 22.3 & $25.55 \pm 0.05$ & $362$ \\
Sextans A & 246.1 & 39.9 & $25.75 \pm 0.1$ & $324 \pm 1$ \\
Sextans B & 233.2 & 43.8 & $25.75 \pm 0.1$ & $300$ \\
M33 & 133.6 & -31.3 & $24.5 \pm 0.1$ & $-179 \pm 3$ \\
\enddata

\tablecomments{
The galaxy sample used for kinematic centre calculations: abbreviated
name, galactic longitude in degrees, galactic latitude in degrees,
distance modulus with uncertainty, heliocentric radial velocity
in km s$^{-1}$ with uncertainty.  All data are
taken from the NED compilation, in some cases with adjustments to
favor more recent and more accurate distance determinations.
Some radial velocity uncertainties were not included in NED and
are not shown.
Galaxies with an asterisk after the name
are closer than 100kpc to either M31 or
the Milky Way; those with two asterisks are within 200kpc.
}
\end{deluxetable}

\clearpage

To perform the correction for Solar motion I follow \citet{PTS11}.
Following the indications in \citet{R09} that the accepted
value of 220 km s$^{-1}$ for the circular velocity of the Sun may
be too small,
I perform calculations for 230 km s$^{-1}$ and separately for
260 km s$^{-1}$, covering the indicated range.  Results for the
230 km s$^{-1}$ correction are designated `w1' hereafter, and
those for 260 km s$^{-1}$ `w2.'
To correct for
Solar motion relative to the Local Standard of Rest I use
the figures of \citet{SBD10}.

\begin{figure}
\plotone{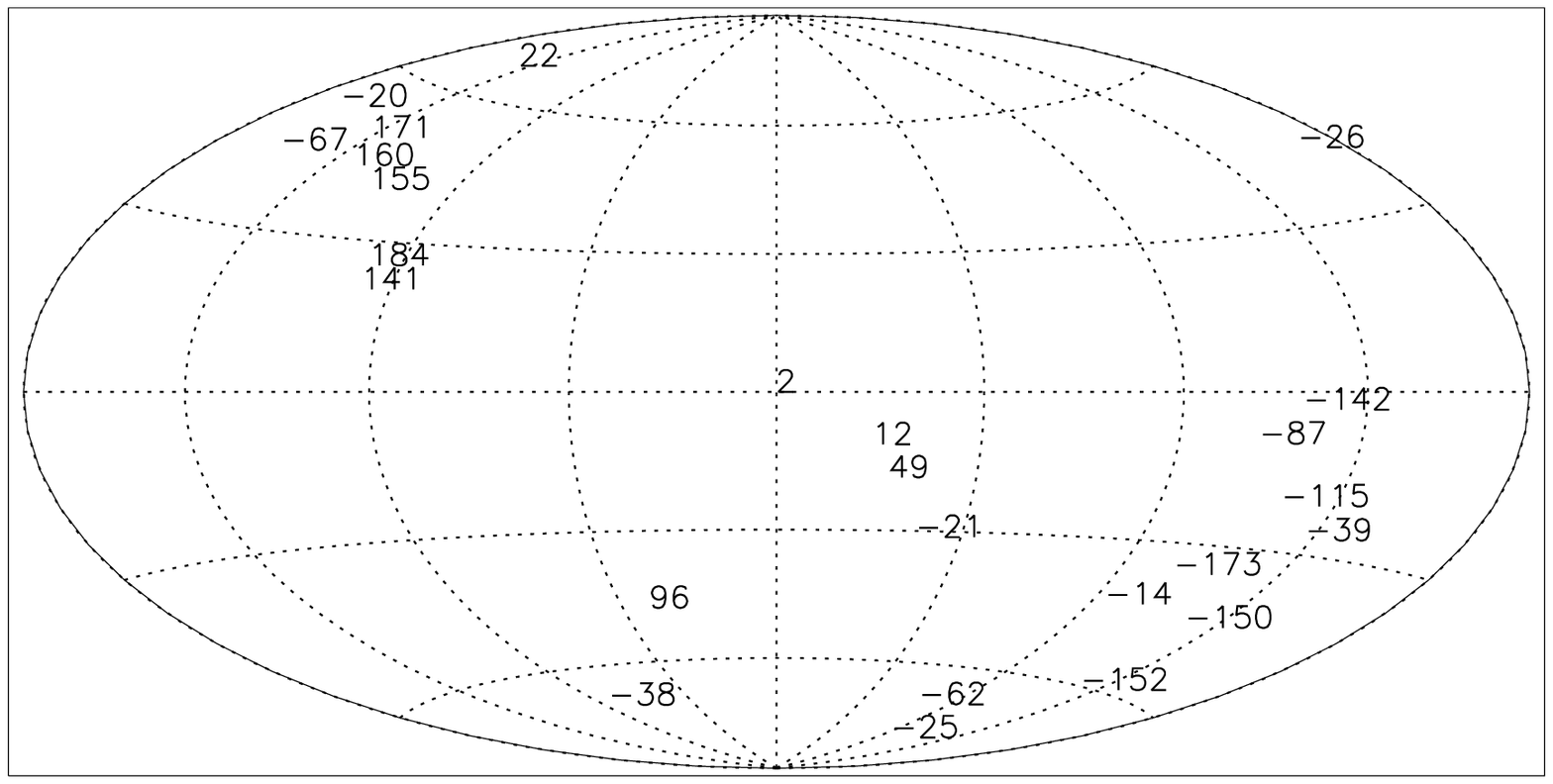}
\caption{The distribution of Local Group radial velocities on the sky,
heliocentric figures corrected for Solar motion using a circular
speed of 230 km s$^{-1}$.  The corresponding plot using a higher
speed gives a qualitatively identical result.}
\label{skyvplot}
\end{figure}

A plot of the (w1) corrected radial velocities on the sky is given
in Fig.~\ref{skyvplot}.  Although there are deviations, in overall
appearance it agrees with the picture of Fig.~\ref{picture}: the
largest magnitudes of radial velocity occur in two roughly opposite
directions, with one of them containing the most negative figures.
A plot of the w2 velocities shows the same pattern.

{\bf The effect of observational uncertainties on our
calculations is expected to be negligible.  Radial velocities are
known to a few km s$^{-1}$, two orders of magnitude more precise
than the Solar motion.  Positions are known to a fraction of a degree;
as will be seen, noise in the velocity field (that is, the fact
that the Local Group does not strictly follow a radial infall pattern)
dominates any error from this source.  Distances are not used in
the following calculation, which seeks only the direction to the center.
They will be used to infer the mass profile, and distance errors
will be considered in that section.}

\subsection{Calculations}

\begin{figure}
\plotone{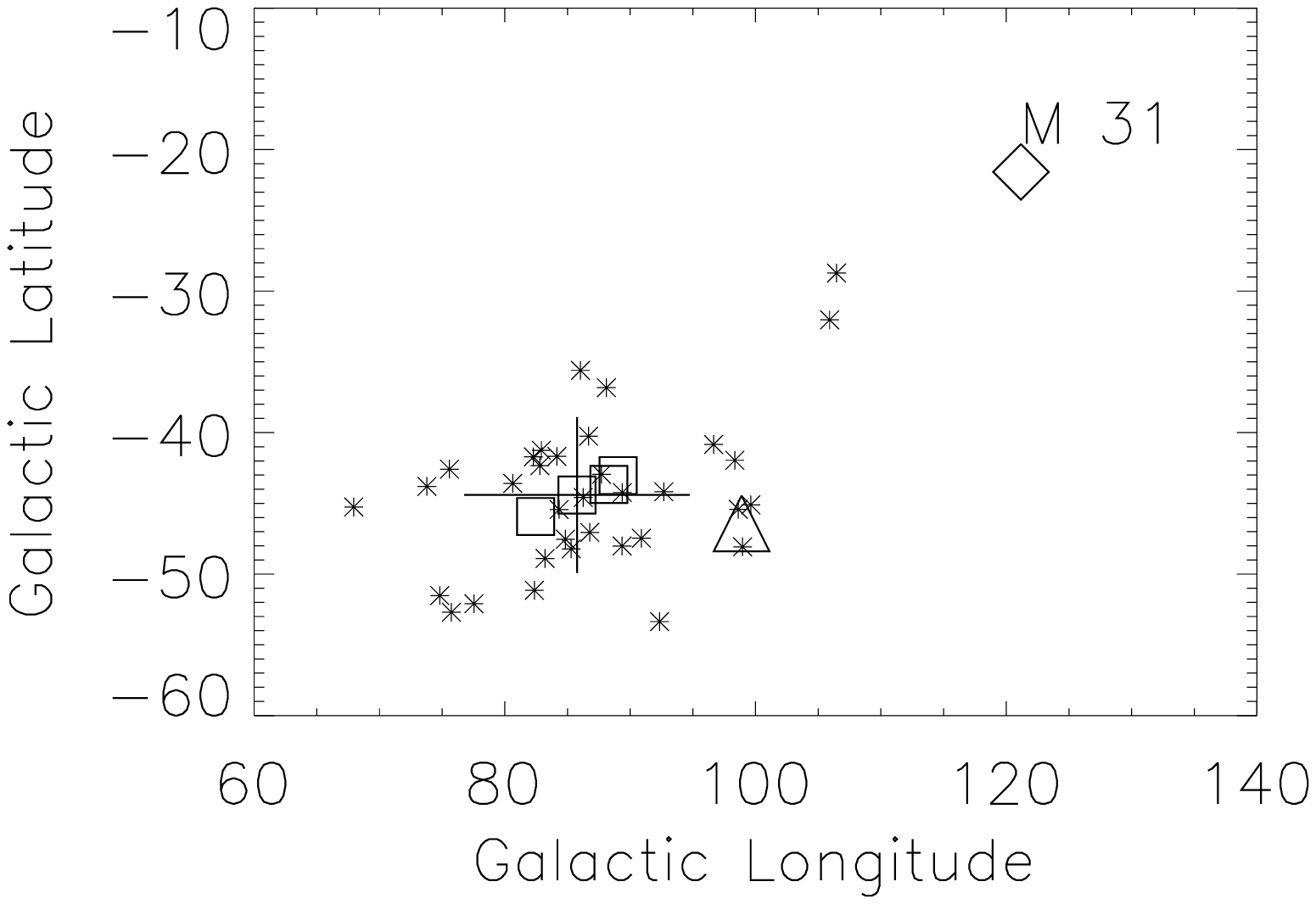}
\caption{The directions of maximum observed radial velocity 
using the $U$-function.  Results for the four different
corrections for Solar motion are shown as squares; that for
the w1 solution has uncertainty estimates attached.  The
jacknife estimate is shown as a triangle and M31 with a
diamond.  Asterisks denote
the various realizations used in the bootstrap calculation.}
\label{vertex1}
\end{figure}

Our goal is to find the direction in which the observed radial
velocities have a maximum in magnitude.  The most straightforward
formulation is to calculate the quantity
\[
U(\hat{\bf{r}}) = \sum_i \left| \hat{\bf{r}} \cdot \bf{v}_i \right|
\]
where $\hat{\bf{r}}$ is the direction we are varying and $\bf{v}_i$
the observed radial velocities, and find the direction that gives
a maximum.  This was done with the w1 and w2 corrections for Solar
motion, and as a check, also with 170 and 200 km s$^{-1}$ corrections.
For the w1 correction a bootstrap calculation yielded uncertainties
in longitude and latitude.  Finally, a jacknife was run to estimate
bias.  The results are given in Table~\ref{directions1} and 
Fig.~\ref{vertex1}.

\begin{table}
\caption{Calculated directions to the Local Group kinematic centre,
using the $U$ function.}
\label{directions1}
\begin{tabular}{@{}lcc}
\tableline
Calculation & Longitude & Latitude \\
w1 correction & 86 $\pm$ 9 & -44 $\pm$ 5.5 \\
w2 correction & 82 & -46 \\
200 km s$^{-1}$ correction & 88 & -44 \\
170 km s$^{-1}$ correction & 89 & -43 \\
jacknife & 99 & -46 \\
\tableline
\end{tabular}

Directions, in Galactic longitude and latitude, for 
the $U$-function
calculations of the Local Group centre.  Uncertainties for
the w1 solution are taken from a bootstrap calculation.
M31 itself lies at 121, -21.
\end{table}

The immediately obvious feature of the results is their offset
from M31, over thirty degrees away. 
It is not due to
an inaccurate correction for Solar motion in the Milky Way;
changing that by some 90 km s$^{-1}$ has very little effect.
The jacknife calculation shows some bias, but nowhere
near enough to explain the offset, and it doesn't lead toward M31.

Note, however, the distribution of our galaxy sample on the sky
in Fig.~\ref{nameplot}; maximising $U$ it is possible we might find some
strange result because the data are sparse and not at all evenly
distributed.
In an effort to correct for this, we divide
by the `shape-function' and instead maximise
\[
V(\hat{\bf{r}}) =\frac{\sum_i \left| \hat{\bf{r}} \cdot \bf{v}_i \right|}
{\sum_i \left| \hat{\bf{r}} \cdot \hat{\bf{r}}_i \right|}.
\]
For various reasons it is actually easier to handle the
related quantity
\[
W(\hat{\bf{r}}) =\frac{\sum_i \left( \hat{\bf{r}} \cdot \bf{v}_i \right)^2}
{\sum_i \left( \hat{\bf{r}} \cdot \hat{\bf{r}}_i \right)^2}
\]
and most calculations will be done with this.

\begin{figure}
\plotone{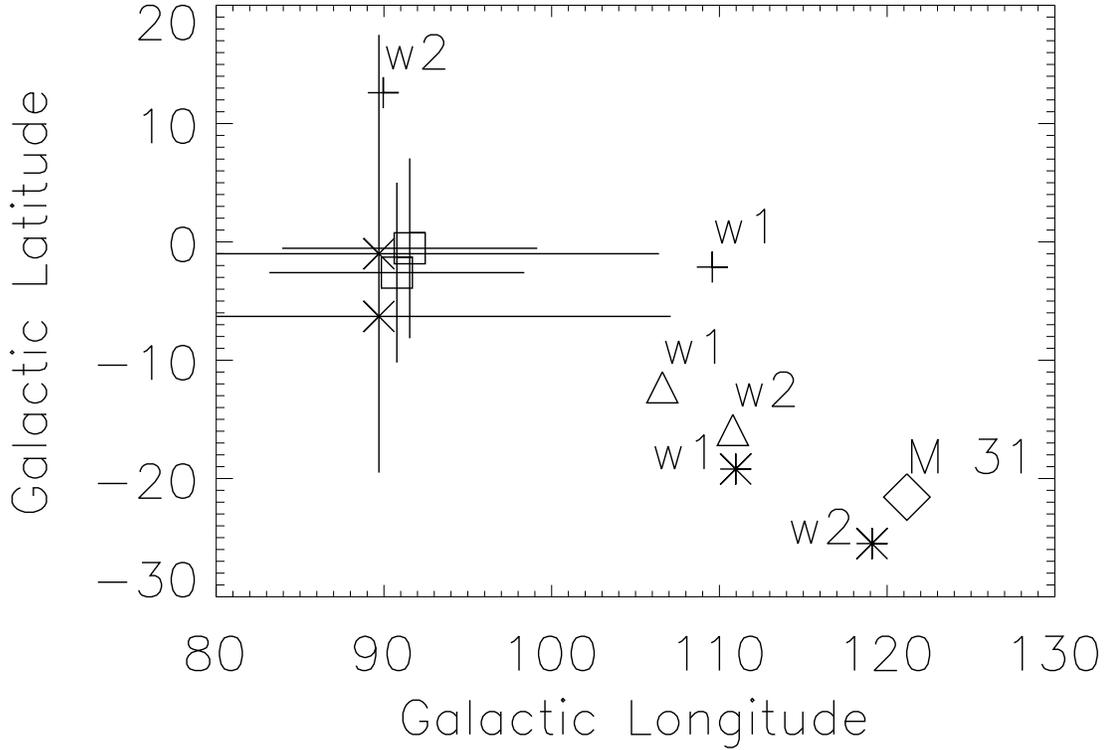}
\caption{The directions of maximum observed radial velocity in the
Local Group under various conditions, along with M31.
The squares show the calculated
vertices for the 25-galaxy sample, with error bars derived from a
bootstrap calculation; the X-symbols with error bars, the average
of 26 run with randomly chosen samples of 20 galaxies;
plus signs, maximising the absolute-value quantity $V$;
triangles,
vertices using the 100 kpc cutoff sample; asterisks, results
using the full sample of 44 galaxies.  Two calculations were done
with each sample, using the two corrections for Solar motion,
labeled `w1' and `w2.'  For the 20- and 25-galaxy calculations,
`w2' is the more northerly result.}
\label{vertex}
\end{figure}

Several variations on the $W$-function calculation were performed.
The the $W$ function was run with four values for
Solar motion, as before, and a jacknife.  
The absolute-value $V$ function was calculated for the w1 and w2
corrections.  In an attempt to
discover the influence of any unusual galaxies or small groups
of them, a bootstrap calculation using only 20-galaxy samples
was included.  Finally, calculations including galaxies
within 200 kpc but outside 100 kpc of the Milky Way and
M31, and then imposing no distance cutoff at all were performed.
The results are tabulated in Table~\ref{directions} and
displayed in Fig.~\ref{vertex}.
(Some of the results do not appear in Fig.~\ref{vertex} for reasons
of clarity.)

The various $V$ and $W$ function calculations show directions much
different from that of the $U$ function, indicating that the shape
of the Local Group has a significant effect on the latter, as we
might expect.  Also as expected, we find that
including the satellite galaxies within 200kpc and then 100kpc
of Andromeda pulls the maximum in that direction.  The bootstrap
calculations again tell us that the offset is real, that the direction
of maximum radial speed is well separated from the line between the
two bright galaxies in the Group.  In addition, the 20-galaxy bootstraps
indicate that this maximum is a feature of the kinematics of the Group
as a whole, not changing greatly when various subsets are excluded.
Corrections for the solar motion make little difference except for
the absolute-value $V$ function.  The jacknife estimate of bias,
however, seems rather wild; it doesn't even fit on the plot of
Fig.~\ref{vertex}.

\begin{table}
\caption{Calculated directions to the Local Group barycentre,
using the `shape' correction.}
\label{directions}
\begin{tabular}{@{}lcc}
\tableline
Calculation & Longitude & Latitude \\
25-galaxy $W$, w1 & 92 $\pm$8 & -0.5 $\pm$ 8 \\
25-galaxy $W$, w2 & 91 $\pm$10 & -2.6 $\pm$ 11 \\
25-galaxy $W$, 200 km s$^{-1}$ & 93 & 0.5 \\
25-galaxy $W$, 170 km s$^{-1}$ & 93 & 2 \\
25-galaxy $V$, w1 & 110 & -2 \\
25-galaxy $V$, w2 & 90 & 13 \\
25-galaxy jacknife $W$, w1 & 108 & 36 \\
20-galaxy average $W$, w1 & 90 $\pm$ 17 & -1 $\pm$ 19 \\
20-galaxy average $W$, w2 & 90 $\pm$ 17 & -6 $\pm$ 22 \\
100-kpc cutoff, w1 & 107 & -12 \\
100-kpc cutoff, w2 & 119 & -26 \\
no cutoff, w1 & 111 & -19 \\
no cutoff, w2 & 119 & -26 \\
\tableline
\end{tabular}

Directions, in Galactic longitude and latitude, for various
calculations of the Local Group barycentre.  The 25-galaxy
sample includes only objects more than 200 kpc from both M31 and
the Milky Way; calculations using a 100kpc cutoff and none at all
are shown for comparison.  M31 itself lies at 121, -21.
\end{table}

\subsection{Asking the right question}

At this point we ask a different question: if the kinematical centre
were located exactly in the Andromeda direction, what would the
calculations come up with?  Of course the various algorithms were tested
on toy velocity models before being employed on actual data, but
these were symmetrical.

\begin{figure}
\plotone{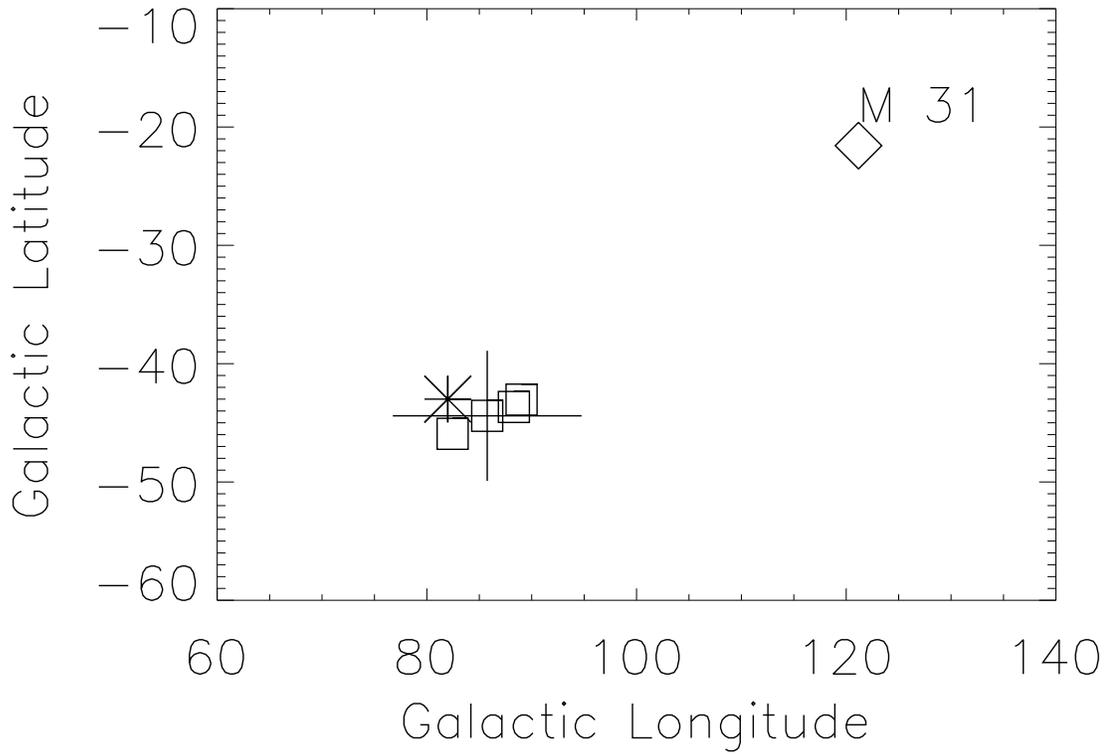}
\caption{The direction of maximum observed radial velocity according to
the $U$ function.  As in Fig.~\ref{vertex1}, 
the squares show the calculation using
observed radial velocities and four different corrections for Solar motion
relative to the Milky Way; the error bars are derived from bootstrap
realizations.  The position of M31 is shown with a diamond.  If all galaxies
are assigned radial velocities that vary as the cosine of angular distance
from Andromeda, the idealized case, the calculated direction is marked
by the asterisk.  The latter is clearly well within the uncertainty of
the data.}
\label{theanswer}
\end{figure}

Now we take the positions of the Local Group galaxies as we find them
on the sky, but assign each a radial velocity equal to 100 km s$^{-1}$
times the cosine of the apparent angle from M31.  The $U$ calculation
now gives a centre in the direction of $l=82, b=-43$, plotted in
figure \ref{theanswer}.  It clearly matches that calculated from
observed radial velocities.

So although the $U$ algorithm works well on symmetrical data, the actual
distribution of Local Group galaxies on the sky forces it to a wrong
answer.  Importantly, it is
one not identified by the jacknife/bootstrap technique.

\begin{figure}
\plotone{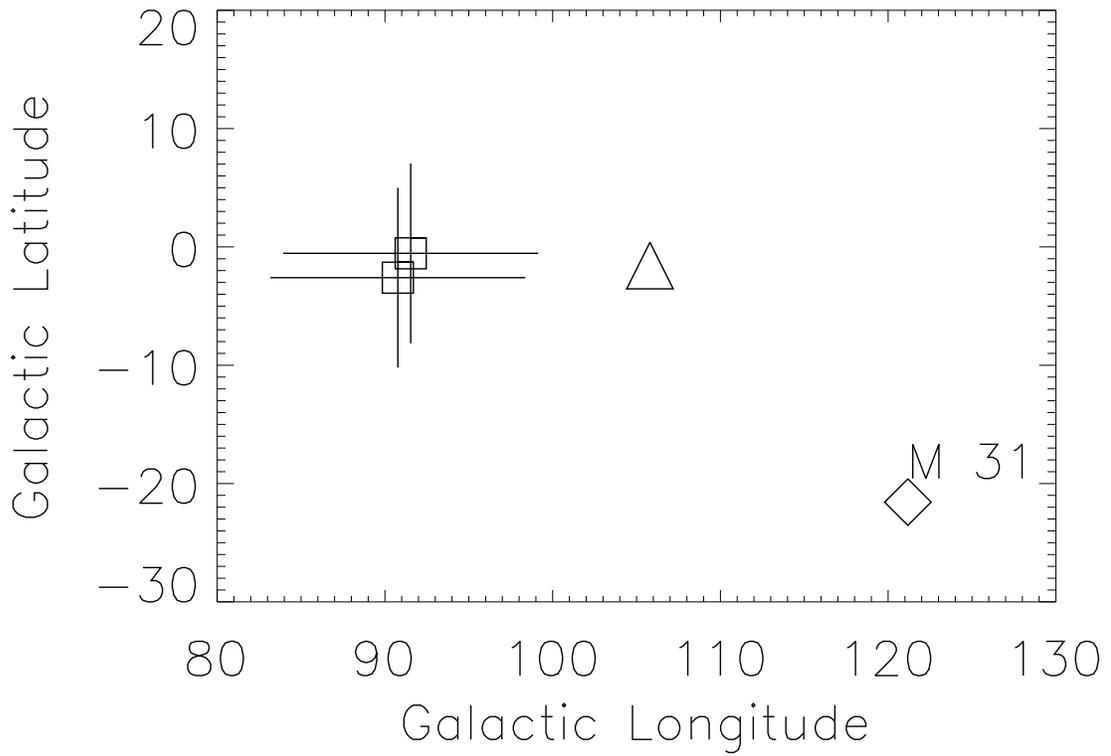}
\caption{The direction of maximum radial velocity according to the
$W$ function.  As in Fig~\ref{vertex1}, 
the boxes show positions based on observed
data, with two different corrections for Solar motion and error bars
derived from bootstrap calculations, and a diamond marks
the position of Andromeda.  The triangle shows the kinematical centre
under the cosine prescription; it remains outside the calculated
uncertainties.}
\label{nottheanswer}
\end{figure}

Applying the same prescription to the $W$ function, we obtain the result
plotted in Fig.~\ref{nottheanswer}.  This time we do not find
such a nice agreement.  The new approach cuts down the offset
greatly but stubbornly remains outside the calculated
error bars.  We should not make too much of these; remember that
the jacknife bias correction does not even fit on this plot.  At any
rate, if
we take this together with the $U$ result, we can conclude with
reasonable certainty that there is no offset.

{\bf Why are the bootstrap/jacknife techniques fooled?  Clearly the problem
lies in the uneven distribution of Local Group galaxies on the sky.  
Beyond that it is difficult to answer.  A detailed look at the statistical
properties of this distribution with reference to standard techniques
would probably be useful, but is beyond the scope of this paper.}

\section{Distance to the center}

\begin{figure}
\plotone{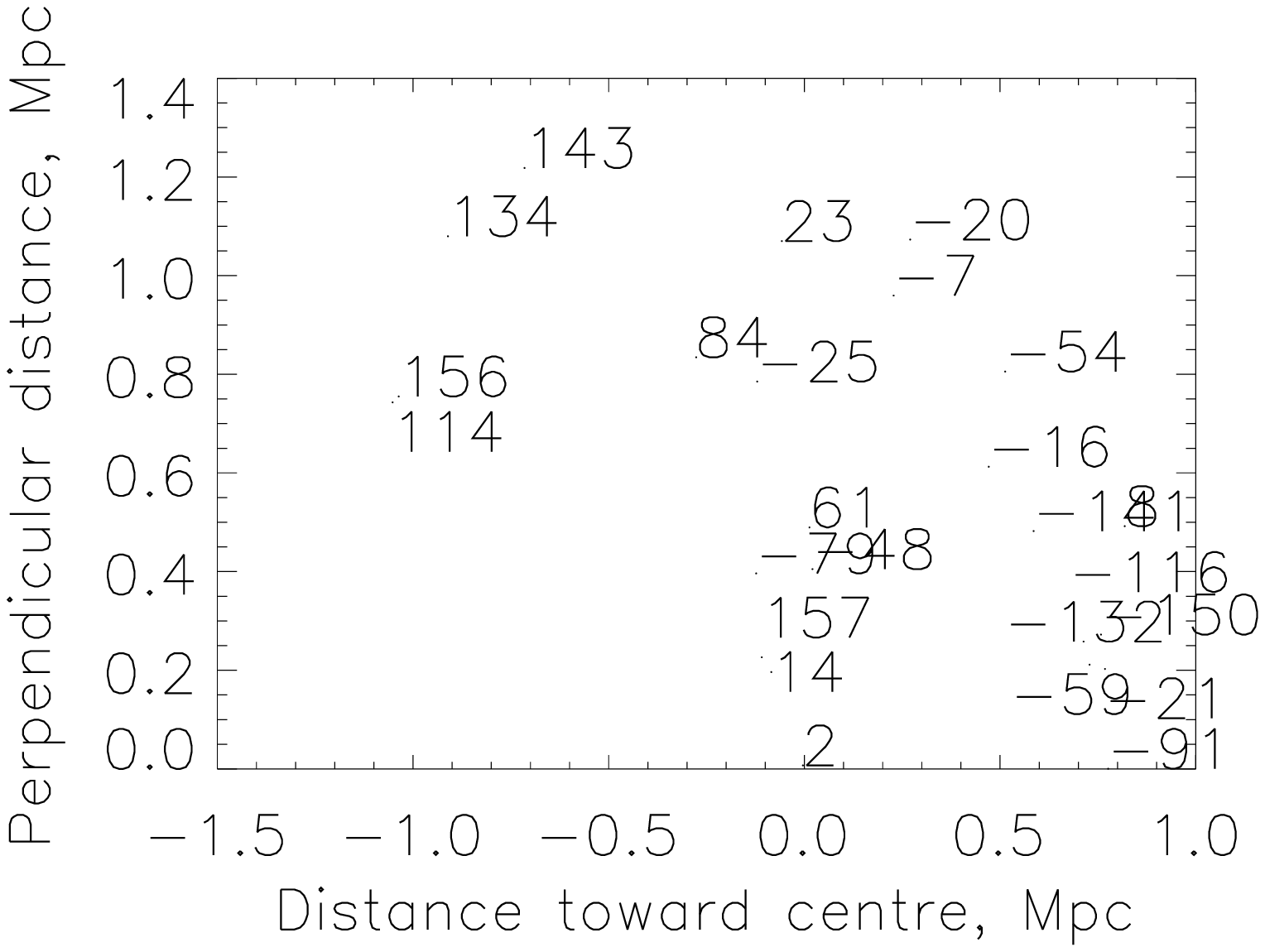}
\caption{Galactocentric radial velocities of Local Group galaxies, plotted
as distance toward the kinematic centre against distance perpendicular
to this line, using the w2 correction for Solar motion.  (A similar plot
for the w1 correction is qualitatively identical, and is used below.)
The absence of galaxies between the Milky Way (at the origin) and M31
(the -91 figure at lower right) shows the centre itself to be unoccupied.
The main feature to notice is the division between negative radial velocities
on the right and positive values on the left.}
\label{w2across}
\end{figure}

\begin{figure}
\plotone{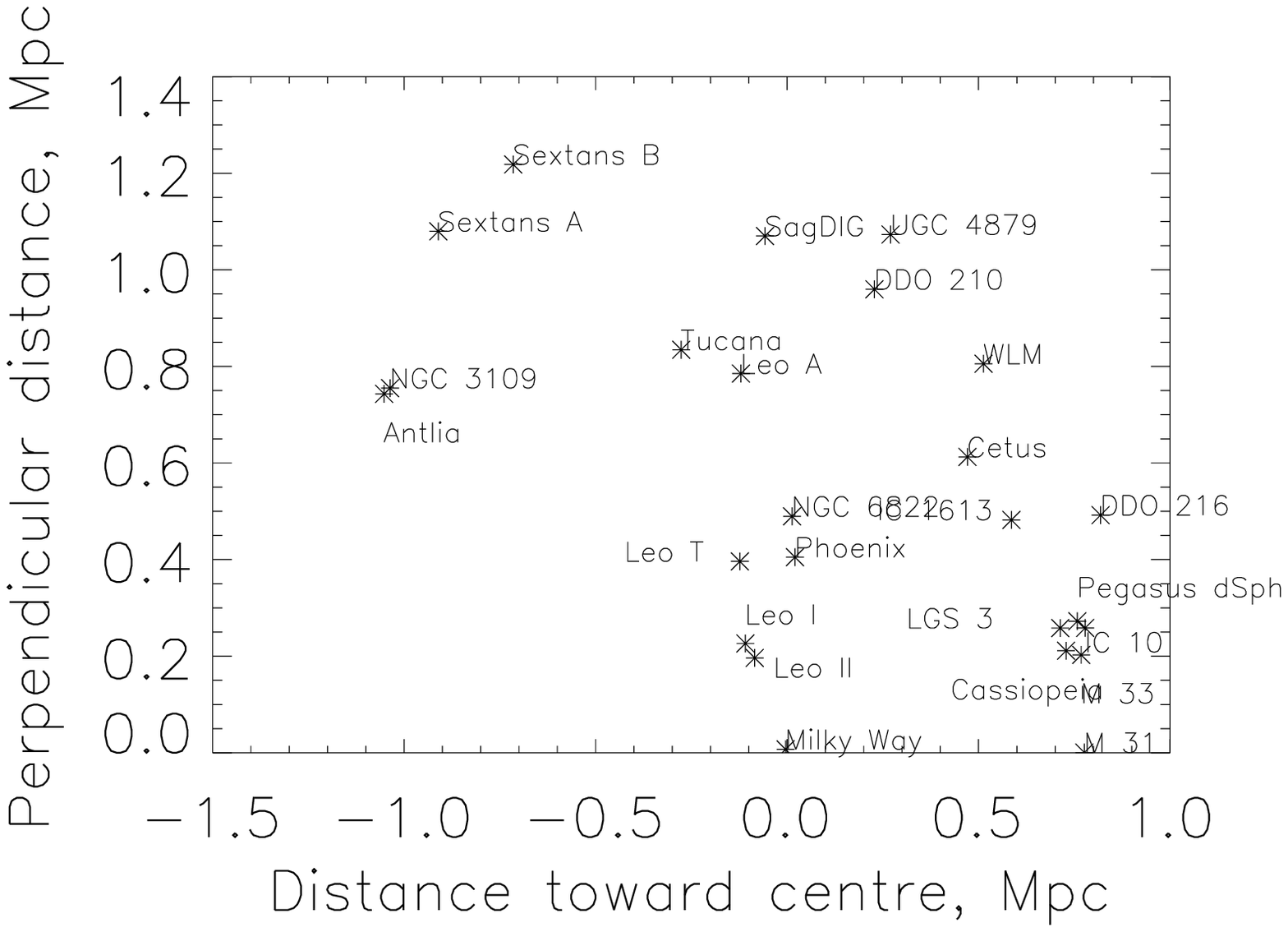}
\caption{Names of the galaxies plotted as distance toward M31 and
perpendicular to that line, as in Fig.~\ref{w2across}.}
\label{nameacross}
\end{figure}

Having satisfied ourselves that the kinematic centre of the Local Group
lies in the direction toward Andromeda, can we contstrain its distance?
This would indeed be useful, possibly showing the relative masses of the
two big spirals.  Keeping in mind Fig.~\ref{picture}, we now plot
galactocentric radial velocities in the direction of M31 and perpendicular
to that line, in Fig.~\ref{w2across}; the galaxies are identified in
Fig.~\ref{nameacross}.

Unfortunately, there are no galaxies between the Milky Way and M31, in
the region where we expect the centre to lie; so there is no direct
indication by a change of sign of radial velocity
of the distance to the centre.  We can only be reasonably
certain that it lies between the generally positive radial velocities
on the left and the negative values on the right\footnote{Working only
from the kinematics, in principle M31 (at -91 km s$^{-1}$ in this figure)
could have fallen through the centre and now be emerging.  This would
require an enormous, completely dark mass, however, somehow located
just along the Milky Way-M31 line and beyond the latter; 
such an implausible situation is
not further considered.}.  To make further headway we have to consider
just how the infall velocities vary with distance.

\section{The velocity-distance law}

Let us designate the position of a galaxy as seen from the centre of the
Group by a vector ${\bf r}$, with ourselves as observers at ${\bf r}_0$.
The angle from ${\bf r}_0$ to ${\bf r}$ at the centre is $\theta$.
A galaxy's motion is radial,  $-a{\bf {\hat r}}$.
The observed radial velocity is then

\begin{equation}
V_{\rm obs} = \frac{- ar - a_or_0 + (ar_0 + a_0r) \cos \theta
}{\sqrt{r^2 + r_0^2 - 2 r r_0 \cos \theta}}
\end{equation}

It would be straightforward to fit the observed
$V_{\rm obs}$ with a function $a$ and a distance to the centre $r_0$
by a standard least-squares or $\chi^2$ technique.  However, we have just
seen how misleading straightforward procedures can be when applied to the
Local Group data.  Instead, we seek simply to match the clearest overall
feature of Fig.~\ref{w2across}, the division between positive and
negative observed radial velocities.  {\bf Note that uncertainties
in distance, of 10\% or less, may blur the picture slightly but
leave this main feature unchanged.  Since the positive-negative division is
roughly radial, distances have little effect on it.}

\begin{figure}
\plotone{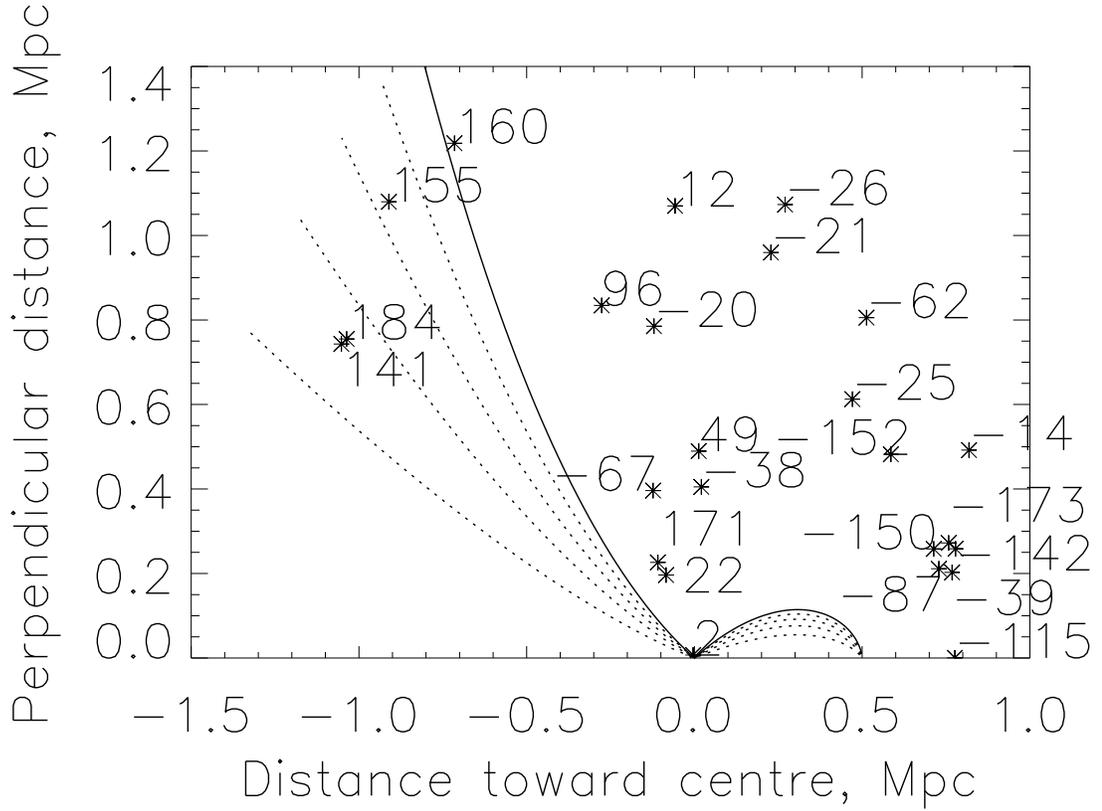}
\caption{Observed radial velocities in the Local Group, as in 
Fig.~\ref{w2across}, but with the w1 correction for Solar motion, and
zero-velocity curves for power-law variations of radial velocity with
distance.  The solid curve shows the expected location of zero observed
radial velocity if $a \propto 1/\sqrt{r}$, as for a single point mass
at the centre of the Group; the dotted curves for the flatter
relations $a \propto r^{-n}$, with $n = 0.1, 0.2, 0.3, 0.4$ left to right.}
\label{powerlaw}
\end{figure}

Setting $V_{\rm obs}=0$ and
dividing through by $a_0 r_0$, we arrive at the condition

\begin{equation}
\alpha \rho + 1 = (\alpha + \rho) \cos \theta
\end{equation}
where $\alpha = a/a_0$ and $\rho = r/r_0$.  In all that follows we take the
distance to the centre as 0.5 Mpc; any reasonable changes make no perceptible
difference to our conclusions.  Taking $a$ as a power law,
$a \propto r^{-n}$, with $n = 0.1, 0.2, 0.3, 0.4, 0.5$ we arrive at the curves
in Fig.~\ref{powerlaw}.

A steeper variation of infall velocity with distance allows a larger 
region of positive observed radial velocity, as one would expect.  For
the steeper law, galaxies closer to the centre are being pulled away from us
more strongly (though no galaxies actually appear here) and we are
being pulled inward more strongly than more distant galaxies.  But note
that the steepest law shown is not strong enough: there are still observed
positive velocities in the negative region, with no negative velocities
in the positive region.  We need to look more closely at our picture
of infall.

\subsection{Falling into a potential well}

We suppose that small galaxies are falling into the general gravitational
potential of the Local Group, trading potential energy for kinetic.
We have
\begin{equation}
\frac{1}{2} v^2(r) - \frac{1}{2} v^2_0 = \Phi_0 - \Phi(r)
\label{potential}
\end{equation}
where $v_0$ is the velocity at some chosen distance where the gravitational
potential is $\Phi_0$.  If the components of the Local Group started their
infall with zero velocity at an infinite distance, or at any rate from so
far away that $\Phi_0$ may be neglected, the infall velocity varies with
distance as
\begin{equation}
v(r) = \sqrt{-2\Phi(r)}
\end{equation}
which, with the point-mass potential of
$\Phi = -GM/r$, gives the $r^{-1/2}$ curve that is not quite
good enough.

Of course galaxies are not point masses, and are indeed generally taken to be
embedded in dark matter haloes.  We will look at two representations
of a large class of halo profiles (\citet{BT08}, pp. 71-2).  The
NFW profile (\citet{NFW96}) is based on n-body simulations
and has a potential
\begin{equation}
\Phi \propto -\frac{\ln \left(1+r/a\right)}{r/a} 
\end{equation}
with $a$ a parameter we will call the core radius, while
that suggested by \citet{LBLB95} was intended to reproduce a flat
rotation curve and has the potential
\begin{equation}
\Phi \propto \ln \left| \frac{\sqrt{s^2+1}+1}{s} \right|
\end{equation}
with $s=r/a$, $a$ again being a parameter we will call the core
radius.

\begin{figure}
\plotone{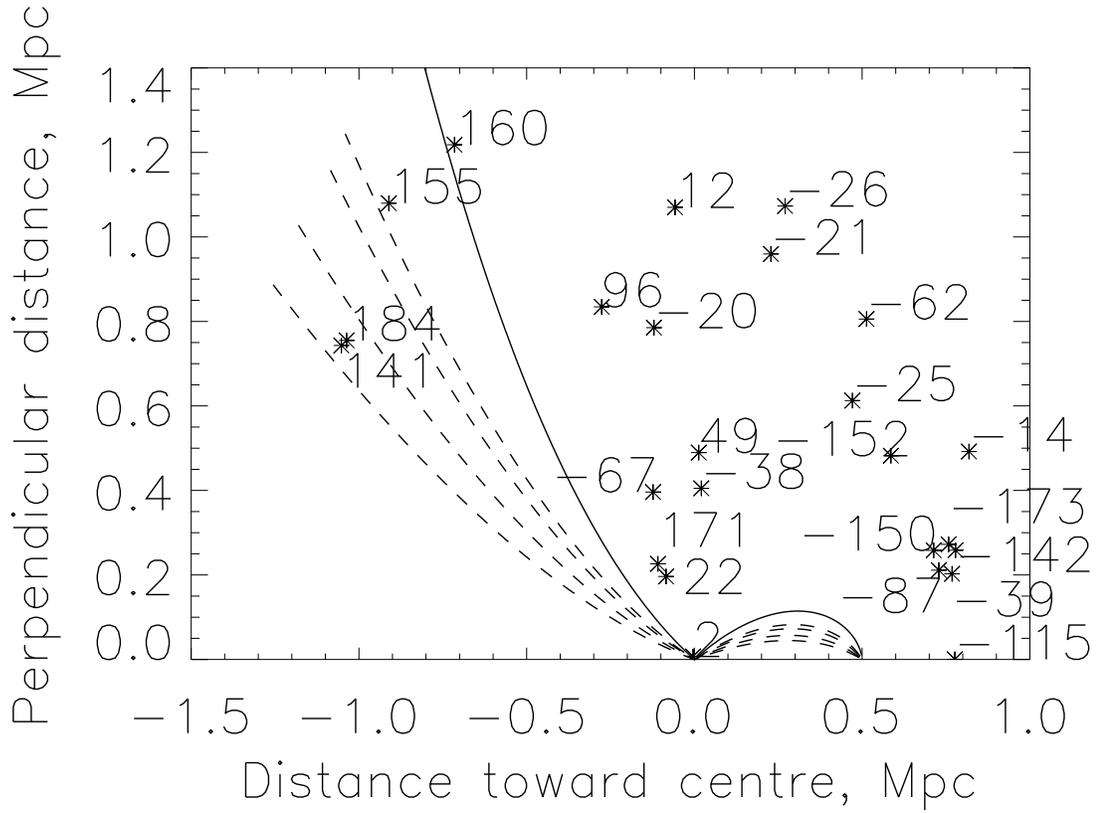}
\caption{Observed radial velocities in the Local Group, as before, with
the $r^{-1/2}$ zero-velocity curve shown solid, and corresponding
curves for NFW halo profiles dotted.  The halo radii for the latter are
0.2, 0.4, 1.0 and 2.0 Mpc, right to left.}
\label{nfw}
\end{figure}

\begin{figure}
\plotone{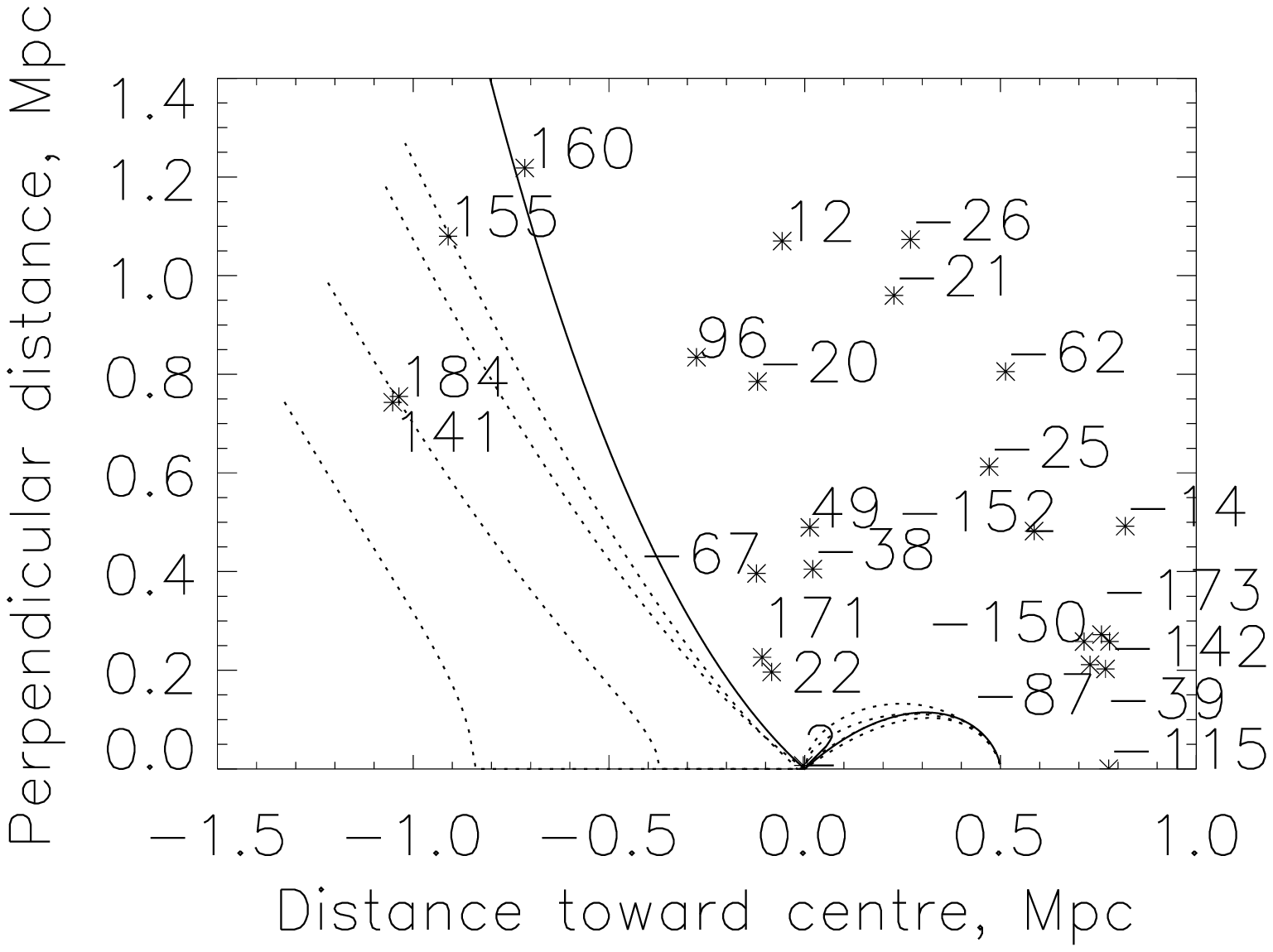}
\caption{Observed radial velocities in the Local Group compared with
zero-velocity predictions for a Lynden-Bell halo model of core radii
0.2, 0.4, 1.0 and 2.0 Mpc (dotted) and the $1/\sqrt{r}$ law (solid).}
\label{lb}
\end{figure}

NFW profiles, with core radii of 0.2, 0.4,
1.0 and 2.0 Mpc gives the curves of Fig.~\ref{nfw} (the $1/\sqrt{r}$
is the solid curve, shown for reference).  For  the Lynden-Bell \&
Lynden-Bell profile
we have the corresponding curves of Fig.~\ref{lb}.

It was not to be expected that spreading out the mass of the Local Group
would give a steeper velocity law, and indeed none of the curves does
as well as  our $r^{-1/2}$.  But 
we have shown that there is no general
dark matter halo around the Local Group, as assumed by
\citet{CL08}, nor indeed any significant
amount of mass apart from the M31-Milky Way pair.  Note that any halo
with $a$ smaller than about 100kpc is indistinguishable from a point
mass by this analysis, and so we cannot say anything about the mass
distribution this close to the centre.

Trying some exotic dynamics, one can generate a promising-looking 
curve by pairing a point mass with a linear repulsion term.  Unfortunately,
to show a good fit to the galactocentric radial velocity data
the repulsion term must be orders of magnitude larger than the observed
cosmological constant or dark energy.  Modified Newtonian Dynamics
(MOND, \citet{M83}) does not fit into the picture at all, since it has
a logarithmic potential (and thus one cannot have galaxies falling in
from infinity).

Several obvious ways to relax our simplifying assumptions do not help.
We could allow galaxies to have different total energies, that is,
different $v=0$ radii. However, that would only flatten the
$v(r)$ function, as closer galaxies would fall in from smaller radii and
thus have lower velocities than before.  It might be possible to arrange
things so that the closer galaxies have systematically
fallen in from farther away,
overtaking the slower ones; but this is contrived and seems unlikely.

We could allow some angular momentum, so that galaxies do not fall
along strictly radial orbits.  Indeed, \citet{B05} found that, in his
N-body simulations that galaxies crossing
the virial radius of a dark-matter halo
had tangential speeds similar to radial speeds.
But this also would flatten the $v(r)$ function, as potential energy
is transformed into tangential motion that (unless carefully arranged
to be always away from the Milky Way)
would be unobservable, or show up only as noise.  In any case, the
virial radius of any dark-matter halo is, as we have seen, likely
to be smaller
than this kind of analysis can distinguish.

\begin{figure}
\plotone{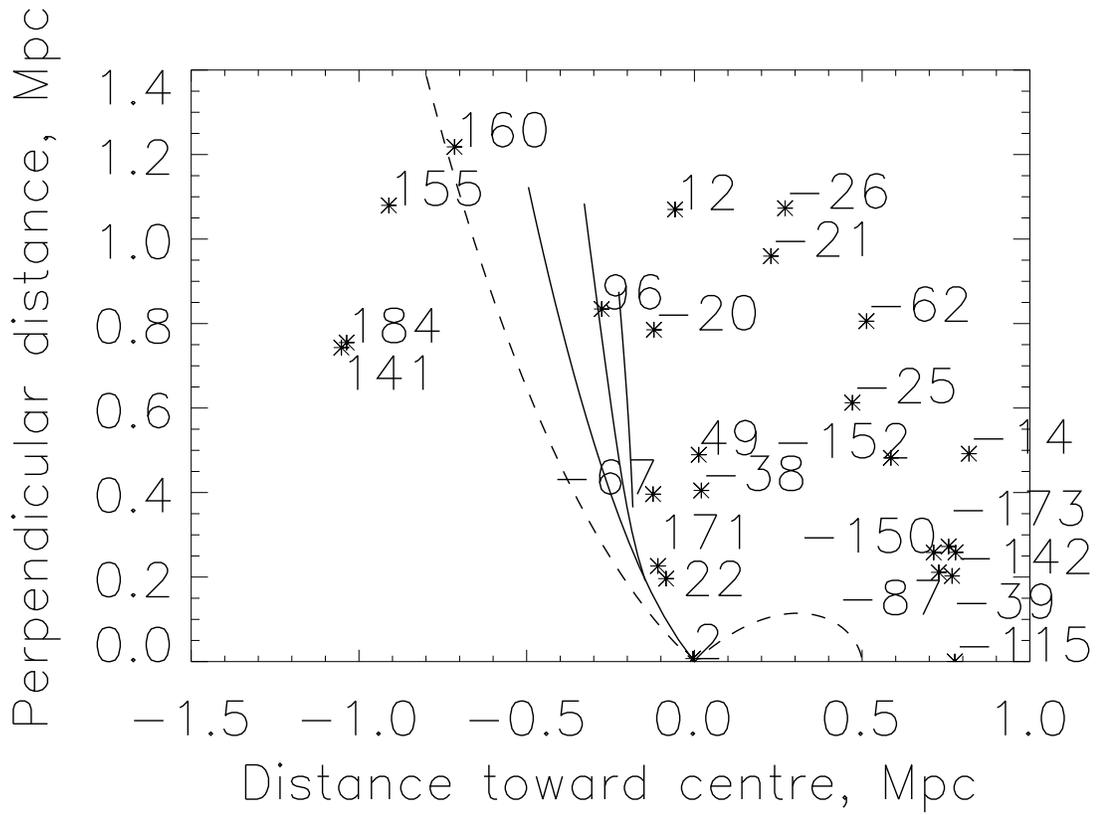}
\caption{Observed radial velocities in the Local Group with
predicted zero-velocity curves.  The dotted curve shows the $r^{-1/2}$
law; the three solid curves include the first multipole term for
two masses separated by 1, 2 and 3 Mpc (left to right).  The separated
masses clearly fit the observations better.}
\label{quadrupole}
\end{figure}

We cannot get a steeper potential well than that of a point mass, clearly.
But by relaxing one of our assumptions we can add higher-order terms to
the potential.  If we have two point masses a distance $b$ apart, the
first two terms in the multipole expansion are

\begin{equation}
\Phi = -\frac{GM}{r} - \mu \frac{GM}{r^3} \frac{b^2 (3 \cos^2 \theta -1)}
{2}
\end{equation}

where $M$ is the total mass, $r$ the distance from the centre of mass
and $\mu$ depends on the mass ratio, being
1/4 for equal masses and 2/9 for a 2:1 ratio.  Using this potential as
we did for the NFW and Lynden-Bell profiles\footnote{The plotted curves assume
a 2:1 mass ratio; equal masses give curves that are indistinguishable.}, 
we get the zero-velocity
curves of Fig.~\ref{quadrupole}.

Two separated masses clearly fit the observations better than any sort
of centered mass profile.  A distance of 1 Mpc between the Milky Way and 
Andromeda is already a significant improvement over a point mass, and
2 Mpc may be a reasonable estimate of the time-averaged distance
between the two galaxies.  3 Mpc fits just a bit better--but at this point
the assumption that $r>b$, upon which the multipole expansion depends, has
completely broken down.  We are at the limit of what our simple picture
can deliver.

\begin{figure}
\plotone{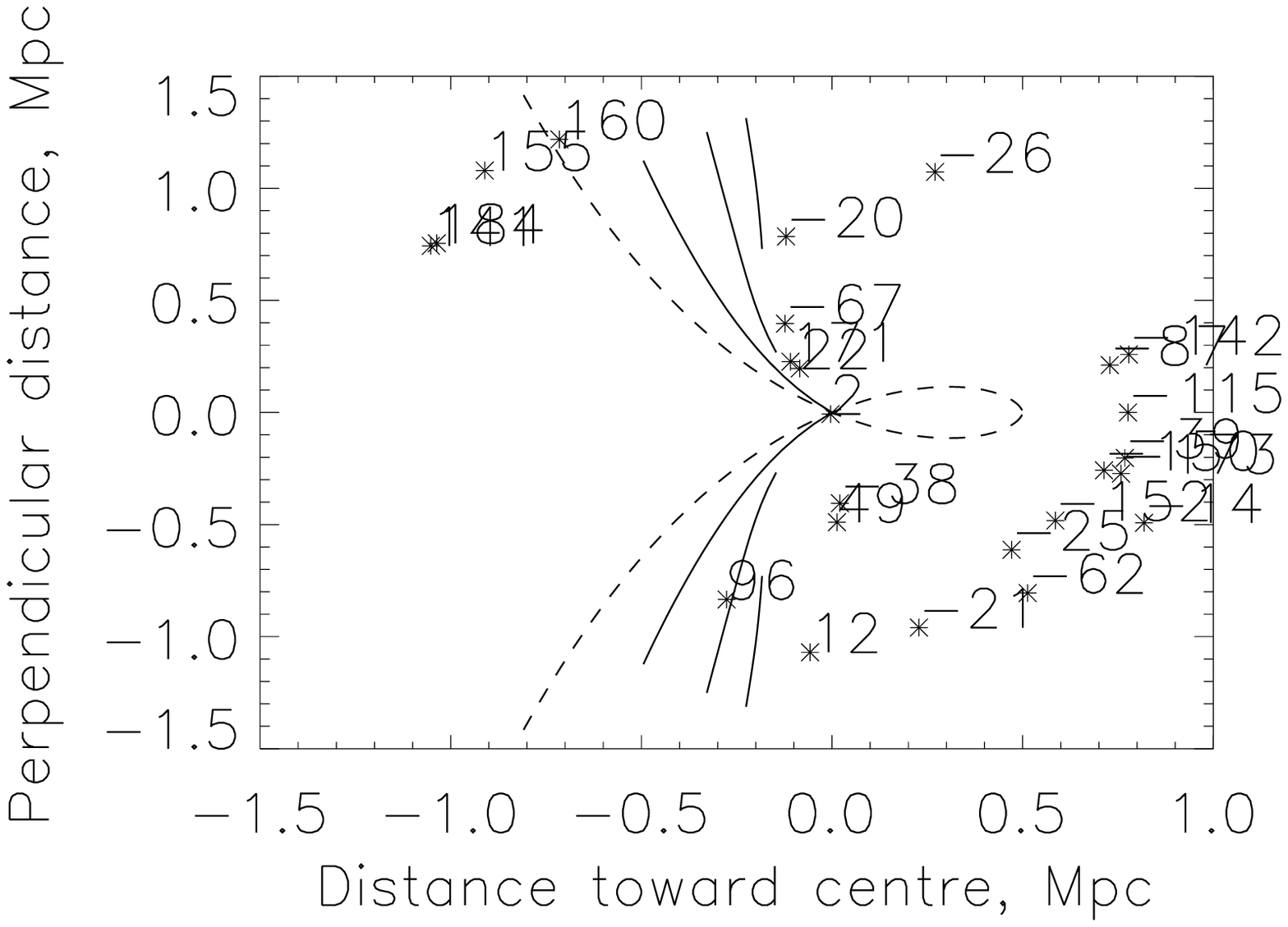}
\caption{The data of Fig.~\ref{quadrupole}, separated into two sides
of the flattened Local Group distribution.}
\label{duplex}
\end{figure}

To check the effect of another of our simplifying assumptions, that
of cylindrical symmetry about the Milky Way-Andromeda axis,
consider the Local Group projected perpendicular
to that axis.  It is rather flattened, stretching almost
2.5 Mpc from Tucana on one side to Sextans A on the other but
less than 1.4 Mpc along a perpendicular line from NGC 3109 to SagDIG.
If we separate the two sides of this very rough plane, like opening a
book in the centre, and plot the information in Fig.~\ref{quadrupole}
again, we arrive at Fig.~\ref{duplex}.

Comparing the two sides, we see that the zero-radial-velocity curve
is somewhat farther to the left on the top than on the bottom.  The
difference, however, is not great, and depends upon one or two
galaxies.  Overall, the symmetry holds up; we are not being misled
by a couple of rogue objects.  But it is also clear that the Local
Group population is sparse in the interesting regions and any further
analysis must be done in a more sophisticated way.

\section{Conclusions}

The primary conclusions of this study are not surprising,
but neither are they trivial.
It has long been assumed that the centre of mass of the Local
Group lies in the direction of M31 and that the Milky Way and
M31 have most of the mass.   
Confirmation of these {\em assumptions} have been arrived at purely
kinematically, using the plausible picture of infall but
without introducing any relations between light and mass at all.
They rule out any extended dark matter halo in the Group as a whole,
as well as any dynamically significant population of orphan haloes.
(They do {\em not} rule out orphan haloes entirely--those are alive and
well among N-body researchers; see, for example, \citet{S14}--but
the Local Group's orphans do not seem to have the dynamic effect that
those in other nearby groups have.)  They agree with
a very recent dynamical study \citep{DKI14}, approaching the problem
in quite a different way.  In addition, it has been shown that the
region from which the Local Group has been assembled is much larger than
its present volume, large enough that the gravitational potential of
most galaxies at the beginning of infall (strictly speaking, the
differences between
them) were negligible.  This could be a very useful result.

An important secondary conclusion is that the kinematics of the Local
Group is {\em not} well-sampled by the visible galaxies.  In fact their
sparseness and asymmetry managed to fool statistical techniques of
moderate sophistication.  The Local Group can be a misleading place!
In turn we reach the conclusion that, 
while things like extended Group haloes and 
dynamically important orphan haloes are clearly worse fits to the data,
it is very difficult to arrive at a convincing numerical estimate of
how much worse they are.

A minor point is that a plausible algorithm (here, a shape correction
term) may not have the desired effect.  In this case, moving from the $U$
to the $V$ and $W$ quantities made the results even more deceiving.

Many years later, the basic simplifying assumptions of \citet{KW59}
have been justified.
Although the sophistication of the dynamical analyses applied to the
Local Group has increased tremendously, the results have been overall a
process of refinement rather than revolution.



\label{lastpage}

\end{document}